\documentclass[a4paper,11pt]{article}
\usepackage[top=30truemm,bottom=30truemm,left=25truemm,right=25truemm]{geometry}
\usepackage{amsmath,amssymb}
\usepackage{amsthm}
\usepackage{ascmac}
\usepackage[hiresbb]{graphicx}
\usepackage{subcaption}
\usepackage{fancyhdr}
\usepackage{float}
\usepackage{here}
\usepackage{url}
\usepackage{color}
\usepackage[colorlinks,citecolor=blue,linkcolor=blue,urlcolor=blue,linktocpage=true]{hyperref}
\usepackage[nosort]{cite}
\usepackage{mathrsfs}
\usepackage{physics}
\usepackage{comment}
\usepackage{dsfont}

\begin{document}

\begin{flushright} 
 DMUS-MP-22/16, 
 KEK-TH-2445,
 YITP-22-118

\end{flushright} 


\vspace{0.1cm}

\begin{center}
  {\LARGE
Partial deconfinement: a brief overview
  }
\end{center}
\vspace{0.1cm}
\vspace{0.1cm}

\begin{center}
Masanori Hanada$^a$ and Hiromasa Watanabe$^{b,c}$
\end{center}

\begin{center}
{\it
$^a$ Department of Mathematics, University of Surrey, Guildford, Surrey, GU2 7XH, UK\\
$^b$ KEK Theory Center, High Energy Accelerator Research Organization,
1-1 Oho, Tsukuba, Ibaraki 305-0801, Japan
\\
$^c$ Yukawa Institute for Theoretical Physics, Kyoto University, Kyoto 606-8502, Japan
}
\end{center}

\vspace{1.5cm}

\begin{center}
  {\bf Abstract}
\end{center}

The confinement/deconfinement transition in gauge theory plays important roles in physics, including the description of thermal phase transitions in the dual gravitational theory.
Partial deconfinement implies an intermediate phase in which color degrees of freedom split into the confined and deconfined sectors.
The partially-deconfined phase is dual to the small black hole that lies between the large black hole and graviton gas. 
Better understandings of partial deconfinement may provide us with a clue how gravity emerges from the field theory degrees of freedom.
In this article, we briefly review the basic properties of partial deconfinement and discuss applications.

\newpage
\section{Brief overview of partial deconfinement}
In large-$N$ gauge theories, the confined and deconfined phases typically have entropy and energy of order $N^0$ and $N^2$, respectively. An immediate question would be: what happens between these two situations, e.g., when the energy is $N^1$ or $\epsilon N^2$ with $\epsilon\ll 1$? The answer is \textit{partial deconfinement}~\cite{Hanada:2016pwv,Berenstein:2018lrm,Hanada:2018zxn,Hanada:2019czd,Hanada:2020uvt,Hanada:2021ipb}. 

\begin{figure}[t]
    \begin{center}
        \includegraphics[width=0.5\textwidth]{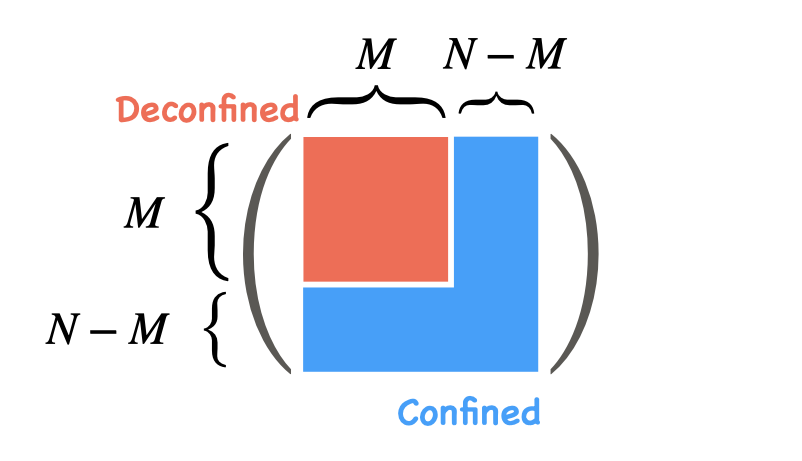}
        \caption{
            Schematic picture of the field configuration that partial deconfinement exhibits.
            The $M\times M$-block (red) and other elements (blue) represent the deconfined and confined sectors, respectively.
            Although this picture shows the adjoint case, we can confirm partial deconfinement in theory with fundamental matters (see Refs.~\cite{Hanada:2019czd,Hanada:2019kue}).
        }
        \label{fig:partial-deconfinement-block}
    \end{center}
\end{figure}

Partial deconfinement can be understood as the coexisting phenomenon in the space of the color degrees of freedom, as depicted in Fig.~\ref{fig:partial-deconfinement-block}. 
The partially-deconfined phase (equivalently, partially-confined phase) is distinguished from the completely-confined phase ($M=0$) and completely-deconfined phase ($M=N$), where $M$ is the ``size'' of the deconfined sector. 
In particular at large $N$, two thermal phase transitions, the Hagedorn transition~\cite{Hagedorn:1965st} ($M=0$ to $M>0$) and the Gross-Witten-Wadia (GWW) transition~\cite{Gross:1980he,Wadia:2012fr} ($M<N$ to $M=N$), clarify the distinction among three phases.

A formal description of partial deconfinement can be obtained by the distribution of Polyakov-loop phases~\cite{Hanada:2020uvt,Hanada:2018zxn}.
Let $A_t$ be the temporal gauge field $A_t$ and define the Polyakov loop by
\begin{equation}
    P 
    \equiv
    \frac{1}{N}\operatorname{Tr}
    \mathcal{P} \exp\left[i\int_0^\beta \mathrm{d} t \,A_t\right]
    = 
    \frac{1}{N} \sum_{j=1}^N e^{i\theta_j}
    =
    \int \mathrm{d}\theta\; \rho^{\rm (P)} (\theta)\, e^{i\theta},
    \label{eq:Polyakov-loop}
\end{equation}
where $\beta$ is the circumference of the temporal direction identified with inverse temperature $\beta = 1/T$, and $\mathcal{P}$ represents the path ordering.
Diagonalizing the matrix before taking trace provides $N$ phases $\theta_j$ lie in $[-\pi, \pi)$ called the Polyakov line phase, 
and in $N\to \infty$ limit, the summation in the middle expression of \eqref{eq:Polyakov-loop} can be replaced by the integration of the phase introducing the distribution function $\rho^{\rm (P)} (\theta)$.
For theories with center symmetry such as pure Yang-Mills theory or maximally supersymmetric Yang-Mills theory, the Polyakov loop is a convenient order parameter for the phase transition associated with the breaking of the center symmetry. 
Recently it has been shown~\cite{Hanada:2020uvt} that the phase transition can be determined from the distribution function $\rho^{\rm (P)} (\theta)$ even for theories without center symmetry, such as QCD.  
In the completely-confined phase, the phases distribute uniformly between $\pm \pi$, i.e., $\rho^{\rm (P)} (\theta)=\frac{1}{2\pi}$. 
In the partially-deconfined phase, they distribute non-uniformly and nonzero everywhere.
At the GWW transition, a gap opens at $\theta=\pm\pi$, 
and the distribution is disjoint at $\pm \pi$ in the completely-deconfined phase.
The minimum value of $\rho^{\rm (P)} (\theta)$ is related to the ``size'' $M$ as $\frac{1}{2\pi}\left(1-\frac{M}{N}\right)$. 


Note that the partial deconfinement can also be understood by the analogous notion of Bose-Einstein condensation (BEC)~\cite{Hanada:2020uvt}.
In the quantum system of many-body identical bosons, an enhancement associated with the redundancy brought by the permutation of particles enables us to express that system in the language of gauge theory.
In this correspondence, the number of degrees of freedom in the confined sector on gauge theory side can be identified with the number of particles in BEC known as the off-diagonal long range order (ODLRO).

Historically, partial deconfinement was proposed for 4d $\mathcal{N}=4$ super Yang-Mills theory as the dual of the small black hole phase~\cite{Hanada:2016pwv}. 
Later, it was argued that partial deconfinement can take place in much simpler theories such as the Gaussian two-matrix model~\cite{Berenstein:2018lrm}. 
In Ref.~\cite{Hanada:2018zxn}, it was conjectured that partial deconfinement is a generic feature associated with the confinement/deconfinement transition, regardless of the order of transition,
and that the GWW transition separates partially- and completely-deconfined phases. 
Explicit demonstrations are provided for several weakly-coupled theories~\cite{Hanada:2019czd,Hanada:2019kue} analytically and for a strongly-coupled matrix model~\cite{Watanabe:2020ufk} via the lattice simulation. Refs.~\cite{Hanada:2020uvt,Hanada:2021ipb} explained the underlying mechanism based on the analogy between confinement at large $N$
and BEC.

\section{Specific examples in gauge theory}
%
The thermodynamics of the weakly-coupled large $N$ gauge theory on $\mathrm{S}^{d-1}$ has been studied in a quite generic manner~\cite{Sundborg:1999ue, Aharony:2003sx, Schnitzer:2004qt}.
The presence of the mysterious intermediate phase, which was later identified with the partially-deconfined phase, had been already recognized at the moment.
As the simplest example capturing their essence, 
let us consider the Gaussian matrix model with gauge-singlet constraint~\cite{Berenstein:2018lrm,Hanada:2019czd,Hanada:2020uvt}.
The Hamiltonian is 
\begin{align}
    \hat{H}
    =
    \frac{1}{2} \sum_I \Tr \qty(
    \hat{P}_I^2+\hat{X}_I^2
    )
    =
    \frac{1}{2} \sum_{I,\alpha} \Tr \qty(
    \hat{P}_{I\alpha}^2+\hat{X}_{I\alpha}^2
    ),
\end{align}
where 
\begin{align}
    \hat{X}_{ij}
    =
    \sum_{\alpha=1}^{N^2-1}\hat{X}_\alpha \tau^\alpha_{ij}, 
    \qquad
        \hat{P}_{ij}
    =
    \sum_{\alpha=1}^{N^2-1}\hat{P}_\alpha \tau^\alpha_{ij}, 
\end{align}
and $N\times N$ matrices $\tau^\alpha$ $(\alpha = 1,\cdots, N^2-1)$ represent the $\mathrm{SU}(N)$ generators normalized as ${\rm Tr}(\tau_\alpha\tau_\beta)=\delta_{\alpha\beta}$. 
The canonical commutation relation 
$
[\hat{X}_{I\alpha}, \hat{P}_{J\beta}] = i\delta_{IJ}\delta_{\alpha\beta}
$
is imposed. 

By the creation and annihilation operators $\hat{A}_{I}^\dagger=\frac{\hat{X}_I-i\hat{P}_I}{\sqrt{2}}$ and $\hat{A}_{I}=\frac{\hat{X}_I+i\hat{P}_I}{\sqrt{2}}$,  
the gauge-invariant Fock states can be written by 
\begin{align}
    \Tr\qty(
    \hat{A}^\dagger_{I}\hat{A}^\dagger_{J}\hat{A}^\dagger_{K}\cdots
    )\ket{0}
    =
    \sum_{i,j,k,l\cdots=1}^N\qty(
    \hat{A}^\dagger_{I,ij}\hat{A}^\dagger_{J,jk}\hat{A}^\dagger_{K,kl}\cdots
    )\ket{0},
    \label{eq:state-SU(N)}
\end{align}
using the Fock vacuum $\ket{0}$ which satisfies
$
    \hat{A}_I \ket{0} = 0
$.
One can also obtain the $\mathrm{SU}(M)$-invariant but $\mathrm{SU}(N)$-noninvariant states 
\begin{align}
    \Tr\qty(
    \hat{A}^\dagger_{I}\hat{A}^\dagger_{J}\hat{A}^\dagger_{K}\cdots
    )\ket{0}
    =
    \sum_{i,j,k,l\cdots=1}^M\qty(
    \hat{A}^\dagger_{I,ij}\hat{A}^\dagger_{J,jk}\hat{A}^\dagger_{K,kl}\cdots
    )\ket{0},
    \label{eq:state-SU(M)}
\end{align}
by restricting the range of the indices.
The difference between \eqref{eq:state-SU(N)} and \eqref{eq:state-SU(M)} is that the indices $i,j,k,l,\cdots$ on the right-hand side runs to $M$ instead of $N$ under restricting an SU($M$)-subsector.
By combining with that type of state, the energy eigenstate preserving $\mathrm{SU}(M)$ symmetry $\ket{E;\mathrm{SU}(M)}$ can be constructed, which is the state corresponding to the field configuration shown in Fig.~\ref{fig:partial-deconfinement-block}.
Although $\ket{E;\mathrm{SU}(M)}$ is not $\mathrm{SU}(N)$-invariant state, the $\mathrm{SU}(N)$-symmetrization 
\begin{equation}
    \ket{E}_\mathrm{inv}
    \equiv
    \frac{1}{\sqrt{\mathrm{Vol(SU} (N) )}} 
    \int_{\mathrm{SU}(N)} \dd U\; \mathcal{U} \qty(\ket{E;\mathrm{SU}(M)})
    \label{eq:SU(N)-symmetrization}
\end{equation}
provides an $\mathrm{SU}(N)$-invariant eigenstate. 
Here $\mathcal{U}$ represents the $\mathrm{SU}(N)$ gauge transformations associated with a group element $U$, and the integration is performed over $\mathrm{SU}(N)$ with the Haar measure. 
By construction, $\ket{E}_\mathrm{inv}$ is SU($N$)-invariant. 
It shows a mapping of the eigenstates from the $\mathrm{SU}(M)$ to $\mathrm{SU}(N)$ theories in a gauge-invariant manner. 
Such $\mathrm{SU}(N)$-invariant states are dominant in the partially-deconfined phase~\cite{Hanada:2019czd}.\footnote{ 
There is a superselection associated with the large-$N$ limit that allows us to use $\ket{E;\mathrm{SU}(M)}$ instead of $\ket{E}_\mathrm{inv}$. 
We can interpret this as a ``spontaneous breaking" of $\mathrm{SU}(N)$ gauge symmetry~\cite{Hanada:2019czd}.
}

In the above example, the meaning of the ``splitting'' of color degrees of freedom is clear: Only the deconfined sector is excited. This is analogous to the system of $N$ non-interacting indistinguishable bosons, in which some particles are excited while the rest falls into the one-particle ground state, i.e., Bose-Einstein condensation (BEC). 
In fact, BEC is the same phenomenon as confinement, when the system of $N$ indistinguishable bosons is regarded as a gauge theory with S$_N$ gauge group.  
Therefore, by mimicking how BEC is generalized to interacting theories, we can generalize partial deconfinement to finite coupling~\cite{Hanada:2020uvt,Hanada:2021ipb}. 

Ref.~\cite{Watanabe:2020ufk} provided an explicit demonstration at strong coupling by studying a matrix model. 
The separation in Figure~\ref{fig:partial-deconfinement-block} was confirmed as a property of lattice configurations with a certain gauge-fixing condition.

\section{Partial deconfinement and holography}
The confinement/deconfinement phase transition is closely related to the geometric transition in the dual gravitational system via the gauge/gravity duality.
The AdS$_5$/CFT$_4$ correspondence~\cite{Maldacena:1997re} is a well-established example of the duality which gives the relation between 4d $\mathcal{N}=4$ super Yang-Mills theory on $\mathrm{S}^3$ and type IIB superstring theory on $\mathrm{AdS}_5 \times \mathrm{S}^5$.
\begin{figure}[t]
    \begin{center}
        \includegraphics[width=0.75\textwidth]{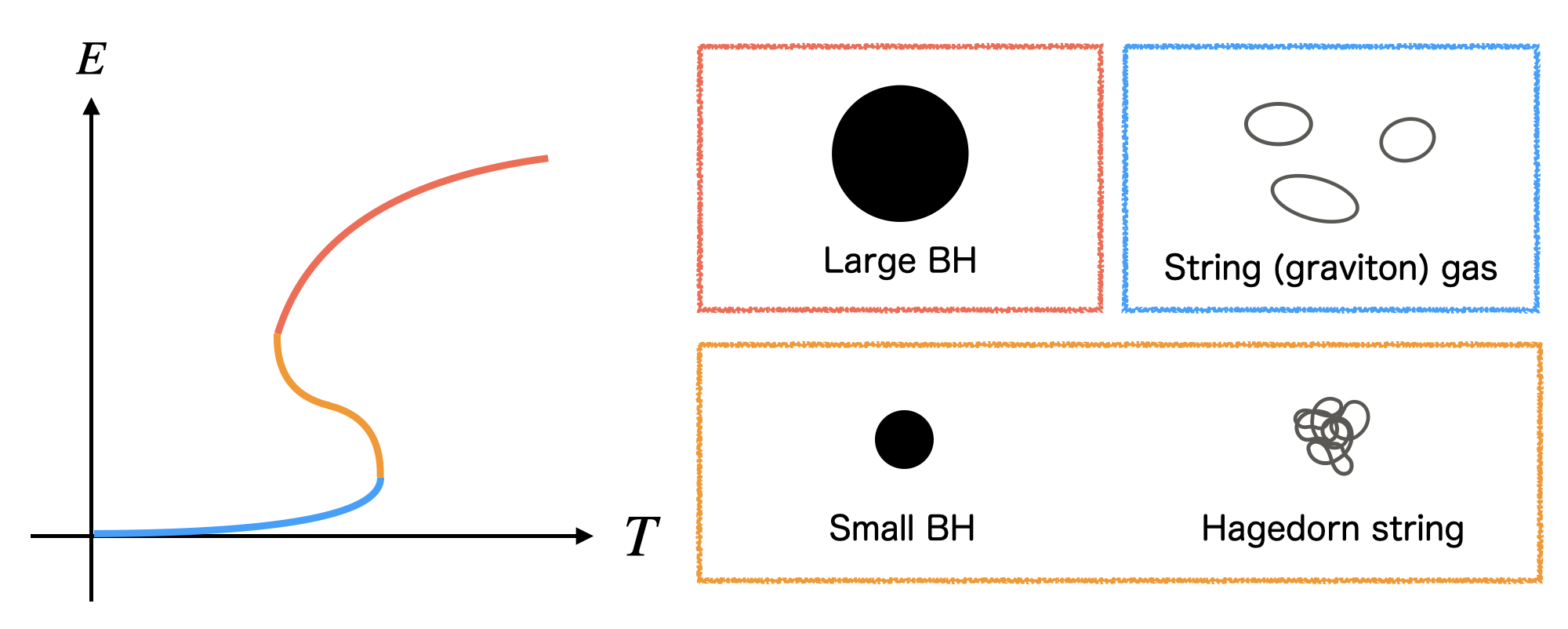}
        \caption{
            Phase structure of the type IIB superstring theory on $\mathrm{AdS}_5 \times \mathrm{S}^5$.
            Same phase structure is expected to be realized for the $\mathcal{N}=4$ SYM on $\mathrm{S}^3$
            at finite temperature
            (See e.g., Ref.~\cite{Aharony:1999ti}).
        }
        \label{fig:phase-structure_gravity}
    \end{center}
\end{figure}
The phase structure had been discussed from both side in the duality~\cite{Aharony:1999ti,Witten:1998zw} (see also Fig.~\ref{fig:phase-structure_gravity});
In the duality, the thermal-AdS phase (i.e., graviton gas phase) and the large black hole phase, which appear as the thermodynamically stable phases, are known as dual to the confined and deconfined phase on the CFT side. These two phases are minima of free energy and correspond to the completely-confined and completely-deconfined phases. 

At the intermediate energy, the ``small'' black hole and the Hagedorn string also exist as thermodynamical states. 
Although these intermediate phases are not favored in the canonical ensemble, they are stable physical states in the microcanonical ensemble~\cite{Aharony:1999ti}. 
The original motivation for considering partial deconfinement was to describe the small black hole in terms of the dual gauge theory~\cite{Hanada:2016pwv,Hanada:2018zxn,Berenstein:2018lrm}.
The small black hole can be described approximately as the 10d Schwarzschild solution.\footnote{Strictly speaking, there are two kinds of small black holes: the 10d Schwarzschild solution and the second branch of the AdS-Schwarzschild solution with the negative specific heat~\cite{Aharony:1999ti}. } 
The energy of such a phase scales as $E\sim N^2T^{-7}$ where $N^2$ can be translated into the Newton constant via the AdS/CFT dictionary, and therefore has the negative specific heat.
It has been a puzzle to identify and express such objects in the dual gauge theory, and after that,
partial deconfinement in gauge theory was proposed as the dual of the small black hole and Hagedorn string.

The supersymmetric matrix models (the BFSS and BMN matrix models~\cite{Banks:1996vh,Berenstein:2002jq}) have been studied intensively in the context of gauge/gravity duality~\cite{Itzhaki:1998dd} as well. These models are expected to describe both string theory and M-theory~\cite{Itzhaki:1998dd,deWit:1988wri,Banks:1996vh}. Of particular interest is how the `transition' between string theory and M-theory takes place. 
The deconfined phase has been studied intensively, and the comparison with the dual type IIA black zero-brane solution was conducted very precisely (see 
the pioneering works~\cite{Anagnostopoulos:2007fw,Catterall:2008yz,Hanada:2008ez,Hanada:2008gy,Catterall:2009xn,Hanada:2013rga,Kadoh:2014hsa,Kadoh:2015mka,Filev:2015hia,Berkowitz:2016jlq,Rinaldi:2017mjl,Asano:2016kxo} and Ref.~\cite{Pateloudis:2022ijr} for the latest result). Recently, the signal of the confined phase in those matrix models at low temperatures was reported~\cite{Bergner:2021goh}. 
The existence of the confined phase and the first-order confinement/deconfinement transition can be natural consequences of dual M-theory description. In this scenario, the partially-deconfined phase corresponds to the 11d Schwarzschild black hole~\cite{Banks:1997hz,Banks:1997tn,Horowitz:1997fr}. 

Moreover, the simple picture adopted above enables us to encode the geometry into color degrees of freedom~\cite{Hanada:2021ipb,Hanada:2021swb};
Very long strings (black hole and Hagedorn string) are dual to the deconfined blocks. In terms of operators, long traces express them. 
Short strings (e.g., gravitons) are short traces corresponding to tiny deconfined blocks. 
Ref.~\cite{Gautam:2022akq} has recently proposed that the black hole evaporation can be visualized in terms of colors, and the Page curve can be derived straightforwardly.

\section{Conclusion and discussion}
The confinement/deconfinement transition is one of the nontrivial features in gauge theory at finite temperature and provides plenty of physical consequences. 
This review focused on partial deconfinement as a key concept and its applications to quantum gravity via holographic duality. 

Another class of important applications is the standard model of particle physics and beyond-standard-model physics.
In the context of the standard model, one possible scenario is that the `cross-over' region of QCD is partially-deconfined with higher-order (i.e., third order) phase transitions. 
It would be interesting if there are experimental signals.
The study of the strongly-coupled lattice gauge theory shows that flux tubes are formed in the confined sector~\cite{Gautam:2022} and the chiral symmetry breaks at the GWW point where the confinement sets in~\cite{Hanada:2021ksu}. 
Although the notion of the GWW transition requires infinite $N$, chiral symmetry breaking is a well-defined concept even at finite $N$. Therefore, we might be able to generalize partial deconfinement to finite $N$ by utilizing chiral symmetry~\cite{Hanada:2021ksu}.
In the context of beyond-standard-model physics, the nature of thermal phase transition may be useful to test the models based on cosmological observations. 

\subsection*{Acknowledgement}
The authors thank the hybrid program ``Nonperturbative and Numerical Approaches to Quantum Gravity, String Theory and Holography'' (code: ICTS/numstrings-2022/9) held in the International Centre for Theoretical Sciences (ICTS) in Bengaluru. The authors were partly supported by ICTS for participating in the program. M.~H. thanks the STFC Ernest Rutherford Grant ST/R003599/1 and the Royal Society International Exchanges award IEC/R3/213026.
H.~W. was supported in part by the JSPS KAKENHI Grant Number JP 21J13014.

\subsubsection*{Data Availability Statement}
No Data associated in the manuscript.

\bibliographystyle{JHEP}
{\footnotesize
\bibliography{ref}

\providecommand{\href}[2]{#2}\begingroup\raggedright\begin{thebibliography}{10}

\bibitem{Hanada:2016pwv}
M.~Hanada and J.~Maltz, \emph{{A proposal of the gauge theory description of
  the small Schwarzschild black hole in AdS$_5\times$S$^5$}},
  \href{https://doi.org/10.1007/JHEP02(2017)012}{\emph{JHEP} {\bfseries 02}
  (2017) 012} [\href{https://arxiv.org/abs/1608.03276}{{\ttfamily
  1608.03276}}].

\bibitem{Berenstein:2018lrm}
D.~Berenstein, \emph{{Submatrix deconfinement and small black holes in AdS}},
  \href{https://doi.org/10.1007/JHEP09(2018)054}{\emph{JHEP} {\bfseries 09}
  (2018) 054} [\href{https://arxiv.org/abs/1806.05729}{{\ttfamily
  1806.05729}}].

\bibitem{Hanada:2018zxn}
M.~Hanada, G.~Ishiki and H.~Watanabe, \emph{{Partial Deconfinement}},
  \href{https://doi.org/10.1007/JHEP03(2019)145}{\emph{JHEP} {\bfseries 03}
  (2019) 145} [\href{https://arxiv.org/abs/1812.05494}{{\ttfamily
  1812.05494}}].

\bibitem{Hanada:2019czd}
M.~Hanada, A.~Jevicki, C.~Peng and N.~Wintergerst, \emph{{Anatomy of
  Deconfinement}}, \href{https://doi.org/10.1007/JHEP12(2019)167}{\emph{JHEP}
  {\bfseries 12} (2019) 167}
  [\href{https://arxiv.org/abs/1909.09118}{{\ttfamily 1909.09118}}].

\bibitem{Hanada:2020uvt}
M.~Hanada, H.~Shimada and N.~Wintergerst, \emph{{Color confinement and
  Bose-Einstein condensation}},
  \href{https://doi.org/10.1007/JHEP08(2021)039}{\emph{JHEP} {\bfseries 08}
  (2021) 039} [\href{https://arxiv.org/abs/2001.10459}{{\ttfamily
  2001.10459}}].

\bibitem{Hanada:2021ipb}
M.~Hanada, \emph{{Bulk geometry in gauge/gravity duality and color degrees of
  freedom}}, \href{https://doi.org/10.1103/PhysRevD.103.106007}{\emph{Phys.
  Rev. D} {\bfseries 103} (2021) 106007}
  [\href{https://arxiv.org/abs/2102.08982}{{\ttfamily 2102.08982}}].

\bibitem{Hagedorn:1965st}
R.~Hagedorn, \emph{{Statistical thermodynamics of strong interactions at
  high-energies}}, {\emph{Nuovo Cim. Suppl.} {\bfseries 3} (1965) 147}.

\bibitem{Gross:1980he}
D.~J. Gross and E.~Witten, \emph{{Possible Third Order Phase Transition in the
  Large N Lattice Gauge Theory}},
  \href{https://doi.org/10.1103/PhysRevD.21.446}{\emph{Phys. Rev.} {\bfseries
  D21} (1980) 446}.

\bibitem{Wadia:2012fr}
S.~R. Wadia, \emph{{A Study of U(N) Lattice Gauge Theory in 2-dimensions}},
  \href{https://arxiv.org/abs/1212.2906}{{\ttfamily 1212.2906}}.

\bibitem{Hanada:2019kue}
M.~Hanada and B.~Robinson, \emph{{Partial-Symmetry-Breaking Phase
  Transitions}}, \href{https://doi.org/10.1103/PhysRevD.102.096013}{\emph{Phys.
  Rev. D} {\bfseries 102} (2020) 096013}
  [\href{https://arxiv.org/abs/1911.06223}{{\ttfamily 1911.06223}}].

\bibitem{Watanabe:2020ufk}
H.~Watanabe, G.~Bergner, N.~Bodendorfer, S.~Shiba~Funai, M.~Hanada, E.~Rinaldi
  et~al., \emph{{Partial deconfinement at strong coupling on the lattice}},
  \href{https://doi.org/10.1007/JHEP02(2021)004}{\emph{JHEP} {\bfseries 02}
  (2021) 004} [\href{https://arxiv.org/abs/2005.04103}{{\ttfamily
  2005.04103}}].

\bibitem{Sundborg:1999ue}
B.~Sundborg, \emph{{The Hagedorn transition, deconfinement and N=4 SYM
  theory}}, \href{https://doi.org/10.1016/S0550-3213(00)00044-4}{\emph{Nucl.
  Phys.} {\bfseries B573} (2000) 349}
  [\href{https://arxiv.org/abs/hep-th/9908001}{{\ttfamily hep-th/9908001}}].

\bibitem{Aharony:2003sx}
O.~Aharony, J.~Marsano, S.~Minwalla, K.~Papadodimas and M.~Van~Raamsdonk,
  \emph{{The Hagedorn - deconfinement phase transition in weakly coupled large
  N gauge theories}},
  \href{https://doi.org/10.4310/ATMP.2004.v8.n4.a1}{\emph{Adv. Theor. Math.
  Phys.} {\bfseries 8} (2004) 603}
  [\href{https://arxiv.org/abs/hep-th/0310285}{{\ttfamily hep-th/0310285}}].

\bibitem{Schnitzer:2004qt}
H.~J. Schnitzer, \emph{{Confinement/deconfinement transition of large N gauge
  theories with N(f) fundamentals: N(f)/N finite}},
  \href{https://doi.org/10.1016/j.nuclphysb.2004.06.057}{\emph{Nucl. Phys. B}
  {\bfseries 695} (2004) 267}
  [\href{https://arxiv.org/abs/hep-th/0402219}{{\ttfamily hep-th/0402219}}].

\bibitem{Maldacena:1997re}
J.~M. Maldacena, \emph{{The Large N limit of superconformal field theories and
  supergravity}}, \href{https://doi.org/10.1023/A:1026654312961,
  10.4310/ATMP.1998.v2.n2.a1}{\emph{Int. J. Theor. Phys.} {\bfseries 38} (1999)
  1113} [\href{https://arxiv.org/abs/hep-th/9711200}{{\ttfamily
  hep-th/9711200}}].

\bibitem{Aharony:1999ti}
O.~Aharony, S.~S. Gubser, J.~M. Maldacena, H.~Ooguri and Y.~Oz, \emph{{Large N
  field theories, string theory and gravity}},
  \href{https://doi.org/10.1016/S0370-1573(99)00083-6}{\emph{Phys. Rept.}
  {\bfseries 323} (2000) 183}
  [\href{https://arxiv.org/abs/hep-th/9905111}{{\ttfamily hep-th/9905111}}].

\bibitem{Witten:1998zw}
E.~Witten, \emph{{Anti-de Sitter space, thermal phase transition, and
  confinement in gauge theories}},
  \href{https://doi.org/10.4310/ATMP.1998.v2.n3.a3}{\emph{Adv. Theor. Math.
  Phys.} {\bfseries 2} (1998) 505}
  [\href{https://arxiv.org/abs/hep-th/9803131}{{\ttfamily hep-th/9803131}}].

\bibitem{Banks:1996vh}
T.~Banks, W.~Fischler, S.~H. Shenker and L.~Susskind, \emph{{M theory as a
  matrix model: A Conjecture}},
  \href{https://doi.org/10.1103/PhysRevD.55.5112}{\emph{Phys. Rev.} {\bfseries
  D55} (1997) 5112} [\href{https://arxiv.org/abs/hep-th/9610043}{{\ttfamily
  hep-th/9610043}}].

\bibitem{Berenstein:2002jq}
D.~E. Berenstein, J.~M. Maldacena and H.~S. Nastase, \emph{{Strings in flat
  space and pp waves from N=4 superYang-Mills}},
  \href{https://doi.org/10.1088/1126-6708/2002/04/013}{\emph{JHEP} {\bfseries
  04} (2002) 013} [\href{https://arxiv.org/abs/hep-th/0202021}{{\ttfamily
  hep-th/0202021}}].

\bibitem{Itzhaki:1998dd}
N.~Itzhaki, J.~M. Maldacena, J.~Sonnenschein and S.~Yankielowicz,
  \emph{{Supergravity and the large N limit of theories with sixteen
  supercharges}}, \href{https://doi.org/10.1103/PhysRevD.58.046004}{\emph{Phys.
  Rev.} {\bfseries D58} (1998) 046004}
  [\href{https://arxiv.org/abs/hep-th/9802042}{{\ttfamily hep-th/9802042}}].

\bibitem{deWit:1988wri}
B.~de~Wit, J.~Hoppe and H.~Nicolai, \emph{{On the Quantum Mechanics of
  Supermembranes}},
  \href{https://doi.org/10.1016/0550-3213(88)90116-2}{\emph{Nucl. Phys.}
  {\bfseries B305} (1988) 545}.

\bibitem{Anagnostopoulos:2007fw}
K.~N. Anagnostopoulos, M.~Hanada, J.~Nishimura and S.~Takeuchi, \emph{{Monte
  Carlo studies of supersymmetric matrix quantum mechanics with sixteen
  supercharges at finite temperature}},
  \href{https://doi.org/10.1103/PhysRevLett.100.021601}{\emph{Phys. Rev. Lett.}
  {\bfseries 100} (2008) 021601}
  [\href{https://arxiv.org/abs/0707.4454}{{\ttfamily 0707.4454}}].

\bibitem{Catterall:2008yz}
S.~Catterall and T.~Wiseman, \emph{{Black hole thermodynamics from simulations
  of lattice Yang-Mills theory}},
  \href{https://doi.org/10.1103/PhysRevD.78.041502}{\emph{Phys. Rev.}
  {\bfseries D78} (2008) 041502}
  [\href{https://arxiv.org/abs/0803.4273}{{\ttfamily 0803.4273}}].

\bibitem{Hanada:2008ez}
M.~Hanada, Y.~Hyakutake, J.~Nishimura and S.~Takeuchi, \emph{{Higher derivative
  corrections to black hole thermodynamics from supersymmetric matrix quantum
  mechanics}},
  \href{https://doi.org/10.1103/PhysRevLett.102.191602}{\emph{Phys. Rev. Lett.}
  {\bfseries 102} (2009) 191602}
  [\href{https://arxiv.org/abs/0811.3102}{{\ttfamily 0811.3102}}].

\bibitem{Hanada:2008gy}
M.~Hanada, A.~Miwa, J.~Nishimura and S.~Takeuchi, \emph{{Schwarzschild radius
  from Monte Carlo calculation of the Wilson loop in supersymmetric matrix
  quantum mechanics}},
  \href{https://doi.org/10.1103/PhysRevLett.102.181602}{\emph{Phys. Rev. Lett.}
  {\bfseries 102} (2009) 181602}
  [\href{https://arxiv.org/abs/0811.2081}{{\ttfamily 0811.2081}}].

\bibitem{Catterall:2009xn}
S.~Catterall and T.~Wiseman, \emph{{Extracting black hole physics from the
  lattice}}, \href{https://doi.org/10.1007/JHEP04(2010)077}{\emph{JHEP}
  {\bfseries 04} (2010) 077} [\href{https://arxiv.org/abs/0909.4947}{{\ttfamily
  0909.4947}}].

\bibitem{Hanada:2013rga}
M.~Hanada, Y.~Hyakutake, G.~Ishiki and J.~Nishimura, \emph{{Holographic
  description of quantum black hole on a computer}},
  \href{https://doi.org/10.1126/science.1250122}{\emph{Science} {\bfseries 344}
  (2014) 882} [\href{https://arxiv.org/abs/1311.5607}{{\ttfamily 1311.5607}}].

\bibitem{Kadoh:2014hsa}
D.~Kadoh and S.~Kamata, \emph{{Lattice simulation of lower dimensional SYM with
  sixteen supercharges}}, \href{https://doi.org/10.22323/1.187.0343}{\emph{PoS}
  {\bfseries LATTICE2013} (2014) 343}.

\bibitem{Kadoh:2015mka}
D.~Kadoh and S.~Kamata, \emph{{Gauge/gravity duality and lattice simulations of
  one dimensional SYM with sixteen supercharges}},
  \href{https://arxiv.org/abs/1503.08499}{{\ttfamily 1503.08499}}.

\bibitem{Filev:2015hia}
V.~G. Filev and D.~O'Connor, \emph{{The BFSS model on the lattice}},
  \href{https://doi.org/10.1007/JHEP05(2016)167}{\emph{JHEP} {\bfseries 05}
  (2016) 167} [\href{https://arxiv.org/abs/1506.01366}{{\ttfamily
  1506.01366}}].

\bibitem{Berkowitz:2016jlq}
E.~Berkowitz, E.~Rinaldi, M.~Hanada, G.~Ishiki, S.~Shimasaki and P.~Vranas,
  \emph{{Precision lattice test of the gauge/gravity duality at large-$N$}},
  \href{https://doi.org/10.1103/PhysRevD.94.094501}{\emph{Phys. Rev.}
  {\bfseries D94} (2016) 094501}
  [\href{https://arxiv.org/abs/1606.04951}{{\ttfamily 1606.04951}}].

\bibitem{Rinaldi:2017mjl}
E.~Rinaldi, E.~Berkowitz, M.~Hanada, J.~Maltz and P.~Vranas, \emph{{Toward
  Holographic Reconstruction of Bulk Geometry from Lattice Simulations}},
  \href{https://doi.org/10.1007/JHEP02(2018)042}{\emph{JHEP} {\bfseries 02}
  (2018) 042} [\href{https://arxiv.org/abs/1709.01932}{{\ttfamily
  1709.01932}}].

\bibitem{Asano:2016kxo}
Y.~Asano, V.~G. Filev, S.~Kov\'a\v{c}ik and D.~O'Connor, \emph{{A computer test
  of holographic avour dynamics. Part II}},
  \href{https://doi.org/10.1007/JHEP03(2018)055}{\emph{JHEP} {\bfseries 03}
  (2018) 055} [\href{https://arxiv.org/abs/1612.09281}{{\ttfamily
  1612.09281}}].

\bibitem{Pateloudis:2022ijr}
S.~Pateloudis, G.~Bergner, M.~Hanada, E.~Rinaldi, A.~Sch\"afer, P.~Vranas
  et~al., \emph{{Precision test of gauge/gravity duality in D0-brane matrix
  model at low temperature}},
  \href{https://arxiv.org/abs/2210.04881}{{\ttfamily 2210.04881}}.

\bibitem{Bergner:2021goh}
{\scshape MCSMC} collaboration, \emph{{Confinement/deconfinement transition in
  the D0-brane matrix model -- A signature of M-theory?}},
  \href{https://doi.org/10.1007/JHEP05(2022)096}{\emph{JHEP} {\bfseries 05}
  (2022) 096} [\href{https://arxiv.org/abs/2110.01312}{{\ttfamily
  2110.01312}}].

\bibitem{Banks:1997hz}
T.~Banks, W.~Fischler, I.~R. Klebanov and L.~Susskind, \emph{{Schwarzschild
  black holes from matrix theory}},
  \href{https://doi.org/10.1103/PhysRevLett.80.226}{\emph{Phys. Rev. Lett.}
  {\bfseries 80} (1998) 226}
  [\href{https://arxiv.org/abs/hep-th/9709091}{{\ttfamily hep-th/9709091}}].

\bibitem{Banks:1997tn}
T.~Banks, W.~Fischler, I.~R. Klebanov and L.~Susskind, \emph{{Schwarzschild
  black holes in matrix theory. 2.}},
  \href{https://doi.org/10.1088/1126-6708/1998/01/008}{\emph{JHEP} {\bfseries
  01} (1998) 008} [\href{https://arxiv.org/abs/hep-th/9711005}{{\ttfamily
  hep-th/9711005}}].

\bibitem{Horowitz:1997fr}
G.~T. Horowitz and E.~J. Martinec, \emph{{Comments on black holes in matrix
  theory}}, \href{https://doi.org/10.1103/PhysRevD.57.4935}{\emph{Phys. Rev. D}
  {\bfseries 57} (1998) 4935}
  [\href{https://arxiv.org/abs/hep-th/9710217}{{\ttfamily hep-th/9710217}}].

\bibitem{Hanada:2021swb}
M.~Hanada, \emph{{Large-$N$ limit as a second quantization}},
  \href{https://arxiv.org/abs/2103.15873}{{\ttfamily 2103.15873}}.

\bibitem{Gautam:2022akq}
V.~Gautam, M.~Hanada, A.~Jevicki and C.~Peng, \emph{{Matrix Entanglement}},
  \href{https://arxiv.org/abs/2204.06472}{{\ttfamily 2204.06472}}.

\bibitem{Gautam:2022}
V.~Gautam, M.~Hanada, J.~Holden and E.~Rinaldi, \emph{{Linear confinement in
  the partially-deconfined phase}},
  \href{https://arxiv.org/abs/2208.14402}{{\ttfamily 2208.14402}}.

\bibitem{Hanada:2021ksu}
M.~Hanada, J.~Holden, M.~Knaggs and A.~O'Bannon, \emph{{Global symmetries and
  partial confinement}},
  \href{https://doi.org/10.1007/JHEP03(2022)118}{\emph{JHEP} {\bfseries 03}
  (2022) 118} [\href{https://arxiv.org/abs/2112.11398}{{\ttfamily
  2112.11398}}].

\end{thebibliography}\endgroup
}

\end{document}